\documentclass[a4paper]{article}
\usepackage{epsfig}
\usepackage{amsmath}
\usepackage{amsfonts}
\usepackage{amssymb}
\usepackage{graphicx}
\usepackage{bm}
\usepackage{apalike}
\begin{document}

\title{Maximum memory capacity on neural networks with short-term depression and facilitation}
\author{Jorge F. Mejias and Joaqu\'{\i}n J. Torres \\
Department of Electromagnetism and Matter Physics and \\
Institute ``Carlos I'' for Theoretical and Computational Physics\\
University of Granada, E-18071 Granada, Spain}
\maketitle

\begin{abstract}
In this work we study, analytically and employing Monte Carlo simulations, the influence of the competition between several activity-dependent synaptic processes, such as short-term synaptic facilitation and depression, on the maximum memory storage capacity in a neural network. In contrast with the case of synaptic depression, which drastically reduces the capacity of the network to store and retrieve ``static'' activity patterns, synaptic facilitation enhances the storage capacity in different contexts. In particular, we found {\em optimal} values of the relevant synaptic parameters (such as the neurotransmitter release probability or the characteristic facilitation time constant) for which the storage capacity can be maximal and similar to the one obtained with {\em static} synapses, that is, without activity-dependent processes. We conclude that depressing synapses with a certain level of facilitation allow to recover the good retrieval properties of networks with static synapses while maintaining the nonlinear characteristics of dynamic synapses, convenient for information processing and coding.

\vspace*{-17cm}
\centerline{\Large To appear in Neural Computation, 2009}
\end{abstract}
\vspace*{17cm}

\section{Introduction and model}
One interesting topic which arises in the study of biologically motivated neural systems is how dynamical processes affecting the synapses in different time scales can influence the collective behaviour of large neural assemblies. A well known example is the slow change in the synaptic strength due to a learning process of activity patterns, that can be modelled as a neural network which includes a hebbian prescription for the synaptic intensities~\cite{hebb}. As a consequence of this process,  the network is able to retrieve previously {\em learned} patterns, mimicking the associative memory tasks which occur in the brain \cite{amari72,hopfield,amitAP}. The maximum number of such patterns ($M_{\text{max}}$), per neuron, that the network is able to store without having significant errors in the recovery task is called the critical (or maximum) storage capacity of the system. This capacity is commonly denoted by the ratio $\alpha_c\equiv M_{\text{max}}/N$, with $N$ being the number of neurons in the network. It is well known that the critical storage capacity is affected by certain considerations about real neural systems, such as constraints in the range of values of the synaptic strength \cite{fusi07}, the mean activity level of the stored patterns \cite{tsodyksstorage88,tsodyksstorage91}, or the topology of the network \cite{menzinger,torresNEUCOM}. None of these works take into account, however, that the synaptic strength also varies in short-time scales producing a fluctuating response which can be depressed (synaptic depression) or enhanced (synaptic facilitation) depending on the presynaptic activity \cite{AVSN97,markramPNAS,tsodyksNC}, which defines the so called {\em dynamic synapses.}

Short-term depression and facilitation have been proposed as dynamical processes responsible for many kinds of complex behaviour found in neural systems. For instance, recent theoretical works have reported the importance of these mechanisms in the appearance of periodic and chaotic switching between stored activity patterns~\cite{torresNC,torresNC07}, which could be related to the oscillations between up and down cortical states~\cite{tsodyks06}. They are also responsible for the generation of persistent activity in working memories \cite{mongillo06,tsodyks07,mongillo08}, and enhance the detection of correlated inputs in a background of noisy activity under different conditions \cite{mejiasCD07}. Synaptic depression occurs due to the existence of a {\em limited} amount of neurotransmitter vesicles in the synaptic button, ready to be released into the synaptic cleft if a presynaptic action potential (AP) arrives. This produces, for a high frequency stimulus, a decrease of the postsynaptic response ---which is a measure of the synaptic strength--- as is shown in~\cite{AVSN97}. On the other hand, synaptic facilitation takes into account the effect of the influx of extracellular calcium ions into the neuron near the synapse after the arrival of each presynaptic AP~\cite{bertramJNEURO}. These ions bind to some sensor which favours the neurotransmitter vesicle depletion, in such a way that the postsynaptic response increases for successive APs \cite{zucker94}. Facilitation, therefore, increases the synaptic strength for high frequency presynaptic stimuli. The effect of the competition between these two a priory opposite mechanisms has been shown to be highly relevant in the emergent behaviour of attractor neural networks (ANN) with activity-dependent dynamic synapses with a finite number of stored patterns~\cite{torresNC07}. However, until now the effect of such competition in the critical storage capacity has not been reported. Only very recently, a few studies have analysed this particularly interesting issue for depressing synapses, and showed that the critical storage capacity of stable memory patterns is severely reduced in this case~\cite{bibitchkov,torresCAPACITY,matsumoto07}. Our aim in this work was to compute the critical storage capacity of an ANN, with both depressing and facilitating mechanisms competing in the synapses, to quantitatively analyse the effect of including facilitation in the system. We demonstrate that synaptic facilitation improves the storage capacity with respect to the case of depressing synapses, for a certain range of the synaptic parameters. Moreover, if the level of depression is not very large, facilitation can increase the critical storage capacity, reaching in some cases the value obtained with static synapses ---which is the maximum that one can obtain considering a hebbian learning rule with unbiased random patterns in a fully connected network. Our results suggest that a certain level of facilitation in the synapses might be positive for an efficient memory retrieval, while the function of strongly depressed synapses could be more oriented to other tasks concerning, for instance, the dynamical processing of data.

Our starting point is a fully connected network of $N$ binary neurons whose state ${\bf s}\equiv \{s_i=0,1;\, \forall i=1,\ldots,N\}$ follows a probabilistic {\em Little} (parallel and synchronous) dynamics~\cite{perettoB}:
\begin{equation}
P[s_i(t+1)=1]=\frac{1}{2}\left \{1+\tanh[2\beta (h_i({\bf s},t)-\theta_i)]\right \}\quad \forall i=1\ldots N,
\label{prob}
\end{equation}
where $h_i({\bf s},t)$ is the local field or the total input synaptic current to neuron $i$, namely
\begin{equation}
h_i({\bf s},t)= \sum_{j\neq i} \omega_{ij} x_j(t) F_j(t) s_j(t).
\label{lfp}
\end{equation}
Here, $\beta=T^{-1}$ is a temperature or noise parameter (i.e., for $\beta \rightarrow \infty$ we have a deterministic dynamics), and $\theta_i$ represents the neuron firing threshold. The coefficients $\omega_{ij}$ are fixed synaptic conductances, consequence of the slow learning process of $M$ memory patterns of activity.  In the following we will choose a hebbian prescription for such learning via the standard covariance rule~\cite{amitcov,tsodyksstorage88}
\begin{equation}
\omega_{ij}=\frac{1}{Nf(1-f)}\sum_{\mu=1}^{M} (\xi_i^{\mu}-f) (\xi_j^{\mu}-f),
\label{hebb}
\end{equation}
where $\left\lbrace \xi_i^{\mu}=0,1;\,i=1\ldots N\right\rbrace $ represents the $M$ stored random patterns with mean activity $\langle \xi^\mu_i\rangle =f=1/2.$ On the other hand, the variables $x_j,~F_j$ appearing in $h_i$ describe the short-term depression and facilitation synaptic mechanisms, respectively. We assume that these variables evolve according to the discrete dynamics~\cite{tsodyksNC,torresCAPACITY}
\begin{equation}
x_j(t+1)=x_j(t)+\frac{1-x_j(t)}{\tau_{rec}}-U_{SE}F_j(t)x_j(t)s_j(t)
\label{x}
\end{equation}
\begin{equation}
u_j(t+1)=u_j(t)+\frac{U_{SE}-u_j(t)}{\tau_{fac}}+U_{SE}(1-u_j(t))s_j(t),
\label{u}
\end{equation}
where $u_j(t)\equiv U_{SE} F_j (t).$ Here, $U_{SE}$ represents the maximum fraction of neurotransmitters which can be released in absence of facilitation each time a presynaptic AP arrives to the synapse, and $\tau_{rec}, \tau_{fac}$ are, respectively, the time constants for depressing and facilitating processes.  The dynamics (\ref{x}-\ref{u}) allows to recover the critical storage capacity of the standard Hopfield model ($\alpha_c\simeq 0.138$)~\cite{amit85} for static synapses, that is, when $x_i=F_i=1,~\forall~i,t$. By a simple inspection of Eqs. (\ref{x}-\ref{u}), this limit corresponds to the case of $\tau_{rec},\tau_{fac} \ll 1$  which makes $x_j$ and $u_j \,\,\forall j$  to quickly reach their maximum values, namely $1$  and $U_{SE}$, and implies $x_j=F_j=1\,\,\forall j,t$\footnote{Note that, more precisely, the static synapse limit is obtained for $\tau_{rec},\tau_{fac}\rightarrow 0,$ but due to the discrete dynamics represented by Eqs. (\ref{x}-\ref{u}) one can have some dynamical instabilities during the simulation of the map for very small time constants. However, a continuous version of the dynamics (\ref{x}-\ref{u}) or considering only steady-state conditions (as we assume in this work) allows to consider without any problem that limit.}. In this limit one has the relation $2[h_i({\bf s},t)-\theta_i]=h_i^{H}({\bf s},t)$  where $h_i^{H}({\bf s},t)$  stands for the local field of the classic Hopfield model with zero threshold, which assumes a $\{-1,+1\}$ code for the neuron states and implies for $\theta_i$ the form
\begin{equation}
\theta_i=\frac{1}{2} \sum_{j\neq i} \omega_{ij}.
\label{threshold}
\end{equation}
Instead, we used in this work the $\{1,0\}$ code because it is more related with biology and allows for a clear separation of the synaptic current $h_i({\bf s},t)$ from the neuron threshold $\theta_i$ and, therefore, it enables one to study the effect of synaptic depression and facilitation alone, without including other adaptive effects related, for instance, with threshold dynamics.
\section{Mean-field analysis}
From the definition of $h_i({\bf s},t)$ and Eqs. (\ref{hebb}) and (\ref{threshold}), we obtain
\begin{equation}
2[h_i({\bf s},t)-\theta_i]=\sum_{\mu} \epsilon_i^{\mu} \overline{m}^{\mu}({\bf s},t)-2 \alpha x_i(t) F_i(t) s_i(t) +\alpha
\label{ht}
\end{equation}
where $\alpha\equiv \frac{M}{N}$, $\epsilon_i^{\mu} \equiv 2\xi_i^{\mu}-1$, and $\overline{m}^{\mu}({\bf s},t)\equiv\frac{1}{N}\sum_j \epsilon_j^{\mu} [2 x_j(t) F_j(t) s_j(t)-1]$.

We assume now that the system reaches some stationary state ($t\rightarrow\infty$) which corresponds to a fixed point of the dynamics. In order to work with the term $x_i(t) F_i(t) s_i(t)$ and to obtain an approximately valid mean field theory, we also assume that the working temperature ($T$) in the system is very small (to avoid as much as possible thermal fluctuations). This hypothesis is reasonable because our goal is to compute maximum storage capacity, so we have to perform the limit $T\rightarrow 0.$ One has then two possible scenarios:
\begin{itemize}
\item[(a)]$T=0$. The state of the system is quenched and it does not present any temporal fluctuations in $s_i.$ Therefore, one can assume that in each site $s_i$ takes a fixed value (namely $s_i^\infty= 1,0$) for all times. We can then evaluate the fixed point in Eqs. (\ref{x}) and (\ref{u}) as a function of $s_i^{\infty}$ and obtain~\cite{matsumoto07}
\begin{equation}
F_i=\frac{1+\tau_{fac} s_i^{\infty}}{1+U_{SE} \;\tau_{fac} \; s_i^{\infty}}~;~~~~~~~~x_i=\frac{1}{1+U_{SE} F_i \;\tau_{rec}\; s_i^{\infty}}.
\label{xu0}
\end{equation}
Taking into account that $s_i^{\infty}$ takes the values $\{ 0,1 \}$, we can simplify the expression for the product $x_i F_i s_i$, leading to
\begin{equation}
x_i F_i s_i=\frac{\gamma'}{1+\gamma \gamma'} s_i^{\infty}
\label{xFs1}
\end{equation}
where $\gamma\equiv U_{SE} \tau_{rec}$ and $\gamma'\equiv \frac{1+\tau_{fac}}{1+U_{SE}\tau_{fac}}$. One can easily check that the static limit (Hopfield model) is obtained again for $\tau_{rec}, \tau_{fac} \rightarrow 0$ which implies $\gamma \rightarrow 0,~\gamma' \rightarrow 1,$ respectively.

\item[(b)]{$T \simeq 0$ ($1\ll\beta<\infty$). For very low temperatures and in the steady state, the typical time interval between thermal fluctuations is very large compared with $\tau_{rec}$ and $\tau_{fac},$ due to the exponential dependency on $\beta$ for the probability to have such fluctuations, so between two consecutive fluctuations the condition (\ref{xFs1}) still holds\footnote{Note, for instance, that if neuron $i$ is in the state $s_i^{\infty}=1$ (one has $h_i>\theta_i$) the probability to fluctuate to the state $1-s_i^{\infty}=0$ is $1-p$ with $p$ given by (\ref{prob}), which gives $(1-p)\sim e^{-2\beta (h_i-\theta_i)}\ll 1.$ Similarly, if $s_i^{\infty}=0$ ($h_i<\theta_i$) the probability to fluctuate to the state $s_i^{\infty}=1$ is $p\sim e^{2\beta(h_i-\theta_i)}$ which is also exponentially small.}. Therefore, averaging (\ref{xFs1}) over all temporal fluctuations of $s_i$ during a large time window $\Delta t \rightarrow \infty$ in the steady state, one has
\begin{equation}
x_i F_i s_i \simeq \frac{\gamma'}{1+\gamma \gamma'} \left\langle s_i \right\rangle_t,
\label{xFs2}
\end{equation}
with $\langle s_i\rangle_t\equiv \lim_{\Delta t\rightarrow \infty} \frac{1}{\Delta t}\sum_{t=t_0}^{t_0+\Delta t}s_i(t).$ Note that $\langle s_i \rangle_t=s_i^\infty$ for $T=0 $ and we recover (\ref{xFs1}), so it is reasonable to assume that the approach (\ref{xFs2}) holds for low (non-zero) temperatures.}
\end{itemize}
In order to compute the critical storage capacity let us consider this second scenario, namely the case of $\beta$ very large and finite. In the limit of $N,M\rightarrow \infty$ with $\alpha=M/N$ finite, one can assume the standard mean-field approach $s_i\approx\langle s_i\rangle,$ which is a good approximation for systems involving long-range interactions as in the case we are considering here, that is, a fully connected network. Under this assumption one has $\langle s_i\rangle_t=\langle s_i\rangle$ and the steady-state condition (\ref{xFs2}) allows to write $\overline{m}^{\mu}({\bf s})\approx \overline{m}^{\mu}\equiv \frac{1}{N}\sum_j \epsilon_j^{\mu} [2 \frac{\gamma'}{1+\gamma \gamma'}  \langle s_j\rangle -1].$ This quantity is related with the usual mean-field overlap function $m^{\mu} \equiv \frac{1}{N} \sum_j \epsilon_j^{\mu} \langle \sigma_j\rangle $ (where $\sigma_j=2 s_j-1$) by the expression
\begin{equation}
\overline{m}^{\mu}=\frac{\gamma'}{1+\gamma \gamma'} m^{\mu}-\left( 1-\frac{\gamma'}{1+\gamma \gamma'}\right) B^{\mu},
\label{two-terms}
\end{equation}
where $B^{\mu} \equiv \frac{1}{N} \sum_j \epsilon_j^{\mu}$ is typically of order ${\cal O}\left( \frac{1}{\sqrt{N}} \right)$ for random unbiased patterns.

Expression (\ref{two-terms}) can be used to calculate the steady state mean-field equations for the system if one assumes that the system reaches a steady state in which the network has a macroscopic overlap with a particular pattern, the so called {\em condensed pattern}, with the remaining $M-1$ being of order ${\cal O}(1/\sqrt{N}).$ In the following and without loss of generality, we choose $\mu=1$ as the condensed pattern.

Using the probability (\ref{prob}) in the steady state it is easy to compute $\langle s_i\rangle$ to obtain, for the mean-field overlap function,
\begin{equation}
m^{\mu}=\frac{1}{N} \sum_i \epsilon_i^{\mu} \tanh \left[ 2 \beta (h_i-\theta_i) \right].
\label{standard-overlap}
\end{equation}
If we neglect the self-energy terms in (\ref{ht}) then $2(h_i-\theta_i) \simeq \sum_{\nu} \epsilon_i^{\nu} \overline{m}^{\nu}.$ Inserting this into Eq. (\ref{standard-overlap}) for $\mu=1$ and using (\ref{two-terms}), the steady state mean-field equation for $m^1 \equiv m$ reads
\begin{equation}
m=\left\langle \left\langle \tanh \left[ \beta \left( \frac{\gamma'}{1+\gamma \gamma'} m +\zeta\right) \right] \right\rangle \right\rangle
\label{mprov}
\end{equation}
where $\left\langle \left\langle \cdot\cdot\cdot \right\rangle \right\rangle$ indicates an average over a distribution ${\cal P}(\zeta)$. Here, $\zeta$ is a gaussian white noise due to the effect of $M-1$ non-condensed patterns, and it is obtained, as we will explain later, taking the limit $N \rightarrow \infty$ in the term $\sum_{\mu \neq 1} \epsilon_i^{1} \epsilon_i^{\mu} \overline{m}^{\mu}$. To derive Eq.~\ref{mprov}, we employed the expression $\overline{m}^{1} \simeq \frac{\gamma'}{1+\gamma \gamma'} m^{1}$ for the condensed pattern, after neglecting the ${\cal O}(1/\sqrt{N})$ contribution in (\ref{two-terms}).
Similarly and following standard techniques, one can compute the {\em spin-glass} order parameter in  our probabilistic approach, that is, $q \equiv \frac{1}{N} \sum_i \tanh^2 \left[ 2 \beta \left( h_i-\theta_i\right) \right]$ (see for instance, \cite{hertzB}) which gives
\begin{equation}
q = \left\langle \left\langle \tanh^2 \left[ \beta \left( \frac{\gamma'}{1+\gamma \gamma'} m+\zeta\right) \right] \right\rangle \right\rangle ,
\label{qprov}
\end{equation}
and the pattern interference parameter $r\equiv \frac{1}{\alpha}\sum_{\mu\neq 1}^M (m^\mu)^2$, which in this limit becomes
\begin{equation}
r=\frac{q}{\left( 1-\beta \frac{\gamma'}{1+\gamma \gamma'} (1-q)\right)^2}.
\label{rprov}
\end{equation}
Equations (\ref{mprov}-\ref{rprov}) for $m,~q$ and $r$ constitute the mean-field solution of the model. However, we must characterize the distribution ${\cal P}(\zeta)$ to have a complete solution. From the calculations of Eqs.~(\ref{mprov}-\ref{rprov}), we can obtain the expression for $\zeta$, after taking the limit $N\rightarrow \infty$ and self-averaging over the distribution of random unbiased patterns, from the variable
\begin{equation}
\zeta_i \equiv \sum_{\mu \neq 1} \epsilon_i^{1} \epsilon_i^{\mu} \overline{m}^{\mu}.
\end{equation}
Considering Eq. (\ref{two-terms}), $\zeta_i$ can be written as a combination of the variables $z_1 \equiv \frac{1}{\sqrt{\alpha r}} \sum_{\mu \neq 1} \epsilon_i^{1} \epsilon_i^{\mu} m^{\mu}$ and $z_2 \equiv  \frac{1}{\sqrt{\alpha}} \sum_{\mu \neq 1} \epsilon_i^{1} \epsilon_i^{\mu} B^{\mu},$ and in the limit of interest explained above, both can be considered as uncorrelated gaussian random variables $\textit{N}[0,1].$ In fact, we have
\begin{equation}
\zeta= \sqrt{\alpha r} \frac{\gamma'}{1+\gamma \gamma'} z_1- \sqrt{\alpha}\left( 1-\frac{\gamma'}{1+\gamma \gamma'} \right) z_2 \equiv C_1 z_1-C_2 z_2.
\end{equation}
For simplicity in the calculations, it is convenient to rewrite $\zeta$ as $\overline{\zeta} \equiv \frac{\zeta}{C_1}$. Since $z_1,~z_2$ are normal-distributed, we can compute the probability distribution ${\cal P}(\overline{\zeta})$ employing standard techniques, that is, ${\cal P}(\overline{\zeta}) = \int \int \delta \left[ \overline{\zeta}-z_1+\frac{C_2}{C_1}z_2\right] p(z_1) p(z_2) dz_1 dz_2
$ where $p(z)$ is the normal distribution $\textit{N}[0,1]$. Computing this integral yields ${\cal P} (\overline{\zeta})=N[0,\sigma^2]$, that is, a gaussian distribution with zero mean and variance $\sigma^2 =1+C_2^2/C_1^2$. This allows to consider
\begin{equation}
\zeta\approx\frac{\gamma'}{1+\gamma \gamma'} \left( \alpha r+\alpha \left( \frac{1+\gamma \gamma'-\gamma'}{\gamma'}\right)^2\right)^{1/2} z
\label{noiseterm}
\end{equation}
where $z$ is a normal-distributed variable $\textit{N}[0,1]$.

Finally, the mean-field equations (after introducing the rescaled inverse of the temperature $\widehat{\beta} \equiv \frac{\gamma'}{1+\gamma \gamma'} \beta$) take the form
\begin{equation}
m=\left\langle \left\langle \tanh \left[ \widehat{\beta} \left( m + z \sqrt{ \alpha r+\alpha \left( \frac{1+\gamma \gamma'-\gamma'}{\gamma'}\right)^2}
\right) \right] \right\rangle \right\rangle
\label{m}
\end{equation}
\begin{equation}
q = \left\langle \left\langle \tanh^2 \left[ \widehat{\beta} \left( m+ z \sqrt{ \alpha r+\alpha \left( \frac{1+\gamma \gamma'-\gamma'}{\gamma'}\right)^2}
\right) \right] \right\rangle \right\rangle
\label{q}
\end{equation}
\begin{equation}
r=\frac{q}{\left( 1-\widehat{\beta} (1-q)\right)^2}.
\label{r}
\end{equation}
The equations (\ref{m}-\ref{r}) constitute the complete mean-field solution of the system for a working temperature near to zero $1\ll\beta<\infty$. It is noticeable that the effect of including synaptic depression competing with facilitation is not a simple rescaling of temperature (marked by the presence of $\hat{\beta}$) compared with the case of the classical static Hopfield model. On the contrary, the dynamics of the synapses affects in a different manner the signal and noise terms produced by the interference of the remaining $(M-1)$ patterns. This becomes evident in the explicit expression of the noise term (\ref{noiseterm}). Our results show that this term has a strong influence on the critical storage capacity when depression and facilitation are present, producing a non-trivial behaviour.

Although Eqs. (\ref{m}-\ref{r}) have been derived assuming $1\ll\beta<\infty,$ one can give some arguments to extend their validity for any $T$ if the system reaches a steady state (for instance, a recall, non-recall or a {\em spin-glass} state). In fact, for relatively high temperatures (and sufficiently low values of $\tau_{rec}$ and $\tau_{fac}$) the dynamics (\ref{x}-\ref{u}) for both $x_i(t)$ and $u_i(t)$ is mainly driven for the fluctuating term which contains $s_i(t),$ instead of the deterministic exponential behaviour with time constants $\tau_{rec}$ and $\tau_{fac}.$ Under this condition a plausible hypothesis is to consider both $x_i(t)$ and $u_j(t)$ as binary variables which follow the probabilistic dynamics of $s_i(t)$ and fluctuate in time between two possible values, namely $x^{(1)} $  ($ u^{(1)}$) when $s_i(t)=1,$ and $x^{(0)}$ ($ u^{(0)}$) when $s_i(t)=0.$ A possible choice (but not the only one) for $x^{(1,0)}~(u^{(1,0)})$ is the two steady state values of $x_i(t)~(u_i(t))$ at $T=0$ ---see expressions in Eq. (\ref{xu0}). This choice implies avoiding any temporal correlations or {\em memory} introduced by $\tau_{rec}$ or $\tau_{fac}$ in the values of $x_{i}(t)$ and $u_{i}(t).$ Considering these assumptions one has\footnote{Note that the effect of facilitation and/or depression in this approach is to change the size of the fluctuation between these two values for $x_i(t)$ and $u_i(t).$ For instance, for $\tau_{rec},\;\tau_{fac}\rightarrow 0$ one has
$x^{(1)}=x^{(0)}=1$ and $u^{(1)}=u^{(0)}=U_{SE},$ so $x_i(t)$ and $u_i(t)$ do not fluctuate.}
\begin{equation}
\begin{array}{lll}
x_i (t)\approx 1+\left( \frac{1}{1+\gamma \gamma '}-1 \right) s_i (t)
\\
F_i (t)=\frac{u_{i}}{U_{SE}}\approx 1+(\gamma '-1) s_i (t),
\end{array}
\end{equation}
which gives again
\begin{equation}
x_i(t)F_i(t)s_i(t)= \frac{\gamma '}{1+\gamma \gamma '} s_i (t) \quad \forall t.
\label{xFs3}
\end{equation}
Note that in (\ref{xFs3}) there is now a dependency on $t$ compared with (\ref{xFs1}). Now computing $\langle s_i(t)\rangle$ in the steady state using (\ref{prob}), as in the standard Hopfield model, one obtains again Eqs. (\ref{m}-\ref{r}) which are, therefore, approximately valid for all the range of temperatures of interest\footnote{Some preliminary results in the limit of $\alpha\rightarrow 0$ have confirmed the validity of (\ref{m}-\ref{r}) for any value of $T<T_c,$ if the system reaches a stable fixed point (data not shown).}. However, this strongly relies on the assumption that a fixed point solution will be reached, and in general this may not be true for relatively large values of $\tau_{rec}$ and $\tau_{fac}$. In this situation, some stationary oscillatory states can emerge as a result of the presence of depression and/or facilitation (see, for instance,~\cite{torresNC,torresNC07}). Concretely, the appearance of these oscillatory states is a consequence of the temporal correlations driven by the deterministic part of the dynamics (\ref{x}-\ref{u}). This deterministic part, which is coupled with the stochastic fluctuations driven by $s_j$, can destabilize the fixed point steady states. As a consequence, the system starts to continuously jump between these metastable states. In this study, since we are interested in computing the maximum storage capacity, and this quantity is evaluated at $T=0$, we will not find these oscillatory solutions and the mean-field theory remains valid.

In general, Eqs. (\ref{m}-\ref{r}) cannot be solved analytically. However, one can still get some information about the critical storage capacity because this is computed for $\beta\rightarrow \infty$ (the zero temperature limit). In this situation one can perform the substitutions $
\int \frac{dz}{\sqrt{2 \pi}} \exp(-z^2/2) \tanh [\widehat{\beta} (az+b)] \simeq {\rm erf} \left( \frac{b}{a \sqrt{2}}\right)$ and
$\int \frac{dz}{\sqrt{2 \pi}} \exp(-z^2/2) \left( 1-\tanh^2[\widehat{\beta} (az+b)]\right)  \simeq \frac{1}{a \widehat{\beta}} \sqrt{\frac{2}{\pi}} \exp\left( -\frac{b^2}{2a^2}\right)
$, which yield
\begin{equation}
m \simeq {\rm erf} (y)
\end{equation}
\begin{equation}
q \simeq 1-\frac{1}{a \widehat{\beta}} \sqrt{\frac{2}{\pi}} \exp (-y^2),
\end{equation}
where
\begin{equation}
y \equiv \frac{b}{a\sqrt{2}} \equiv \frac{m}{ \left( 2 \alpha r+2 \alpha \left( \frac{1+\gamma \gamma'-\gamma'}{\gamma'}\right)^2\right)^{1/2} }
\end{equation}
Employing these approaches together with Eq. (\ref{r}) ones obtains a simplified expression for the complete solution, namely

\begin{equation}
y \left[ \sqrt{2 \alpha \left( 1+\left( \frac{1+\gamma \gamma'-\gamma'}{\gamma'} \right)^2 \right)} + \frac{2}{\sqrt{\pi}} \exp (-y^2) \right] = {\rm erf} (y),
\label{closedequation}
\end{equation}
where we assumed $r\simeq1$ in order to get a closed expression. This assumption works well as an approximation since $r \simeq 1$ in the memory phase.
Eq. (\ref{closedequation}) allows to compute the maximal storage capacity for different synaptic conditions, as we will see below, including the competition between synaptic depression and facilitation.
\begin{figure}[t!]
\centerline{\psfig{file=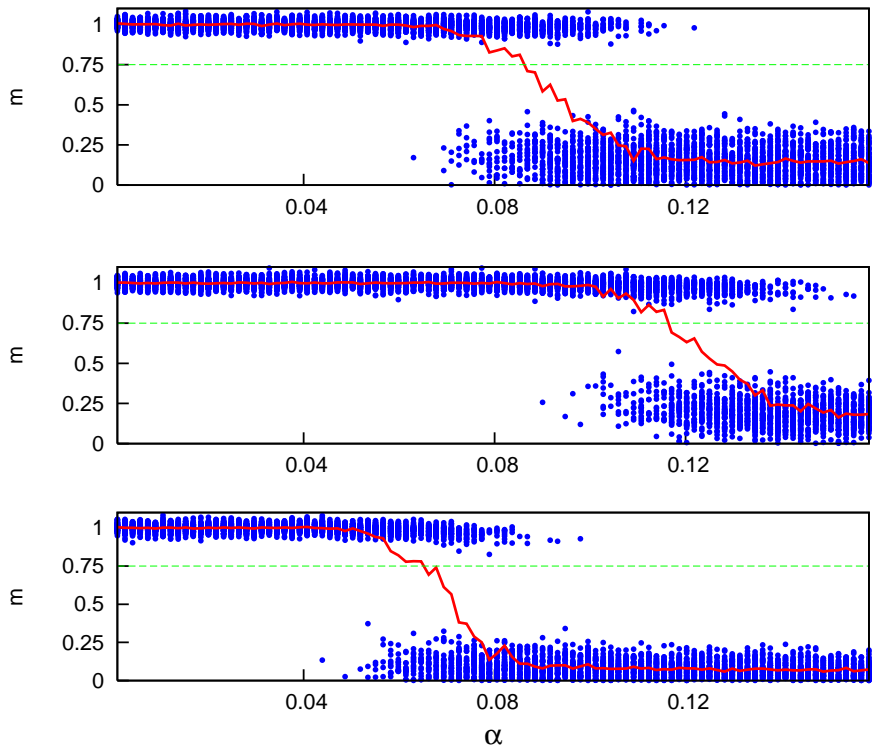,width=12cm}}
\caption{Criterion chosen for the calculation of $\alpha_c$ in Monte Carlo simulations. Each panel shows, for $U_{SE}=0.05,0.5,0.7$ (from top to bottom), a numerical estimate of the critical storage capacity, obtained by averaging the stationary value of the macroscopic overlap (solid line) over many realizations of the stored patterns (dots). The value of $\alpha_c$ corresponds to the crossing point between the averaged overlap and the dashed line $m=0.75$ that we used as criterion for good retrieval of the condensed pattern. Other synaptic parameters were $\tau_{rec}=2$ and $\tau_{fac}=200.$}
\label{fig1}
\end{figure}




\section{Results}
In order to obtain the critical storage capacity $\alpha_c$ for a given set of values of the synaptic parameters, we have to find the maximal value of $\alpha$ for which nontrivial solutions ($y \neq 0$) appear. More specifically, we look for the maximal value of $\alpha$ for which the stationary value of the macroscopic overlap is $m \geq 0.75$. This criterion, which is usually taken for the numerical evaluation of the critical storage capacity in simulations, is illustrated for three different numerical examples in Fig. \ref{fig1}. The figure shows that the value of the critical storage capacity depends on the synaptic parameters, as it was found in \cite{torresCAPACITY,bibitchkov,matsumoto07}. In these works, the inclusion of synaptic dynamics (in particular, synaptic depression) led to a monotonic decrease of the critical storage capacity of the network as one increases the synaptic parameters $\tau_{rec}$ and $U_{SE}.$ This decrease was found to be caused by the loss of stability of the memory fixed points of the system, in the presence of depression. In Fig. \ref{fig1}, however, we see that intermediate values of $U_{SE}$ give higher values of $\alpha_c$, suggesting the possibility of a non-monotonic dependence of $\alpha_c$ on $U_{SE},$ which is mainly due to the presence of facilitation.
\begin{figure}[t!]
\centerline{\psfig{file=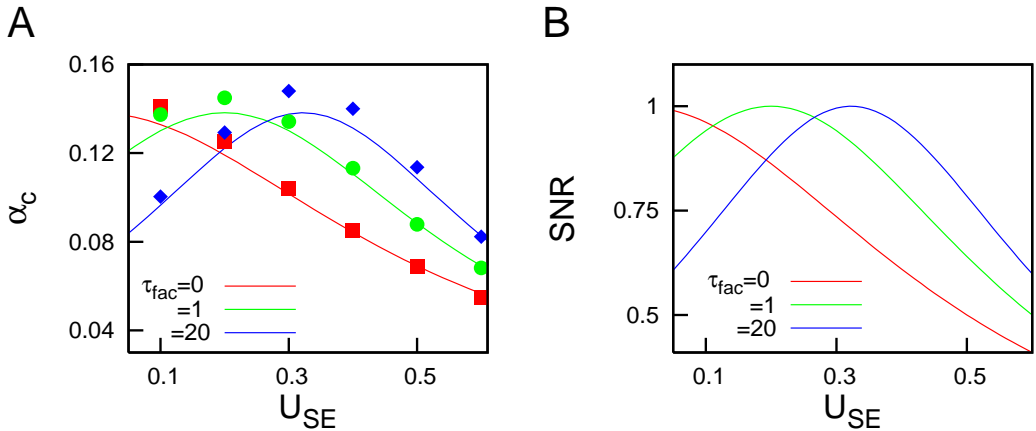,width=14cm}}
\caption{(A) Critical storage capacity $\alpha_c$ as a function of $U_{SE}$ for fixed $\tau_{rec}=2$ and different values of $\tau_{fac}.$ The inclusion of facilitation causes the appearance of a non-monotonic behaviour of the critical capacity of the network as a function of $U_{SE}$ and $\tau_{fac},$ with a maximum which reaches the limit of static synapses ($\alpha_c \simeq 0.138$). Different symbols correspond to numerical simulations of a network with $N=3000$ neurons and different synaptic parameters. (B) Behaviour of $SNR,$ as defined in the text, for the same value of parameters as in panel A. This shows the origin of the non-monotonic behavior of the critical storage capacity found for different values of the synapse parameters.}
\label{fig2}
\end{figure}

A more detailed analysis of this phenomenon is shown in Fig. \ref{fig2}A. The figure shows that the inclusion of synaptic facilitation in a network with depressing synapses ($\tau_{rec}=2$, fixed) induces the appearance of a non-monotonic behaviour of $\alpha_c$ with a maximum value for a given $U_{SE}^*$ which depends on $\tau_{fac}.$ The figure also shows the good agreement of our mean-field theory with simulation of a network of $N=3000$ neurons (symbols). In the particular case of $\tau_{fac}=0,$ we recover the results reported in \cite{torresCAPACITY}, that is, the fact that the static limit $\alpha_c \simeq 0.138$ is only obtained for $\tau_{rec} U_{SE}=0.$ However, if we include the possibility of synaptic facilitation in the synapses, one can obtain $\alpha_c \simeq 0.138$ for $\tau_{rec} U_{SE}>0,$  (that is, for synapses with a certain level of depression). This implies that dynamic synapses are not only convenient for dynamical processing of information in real neurons \cite{AVSN97,abbott04}, or to explain the appearance of global oscillations and other emergent phenomena in neural systems \cite{torresNC07,torresNC,tsodyks06}. In fact, an optimal balance between depression and facilitation is necessary to recover the high retrieval properties of networks with static synapses.

\begin{figure}[t!]
\centerline{\psfig{file=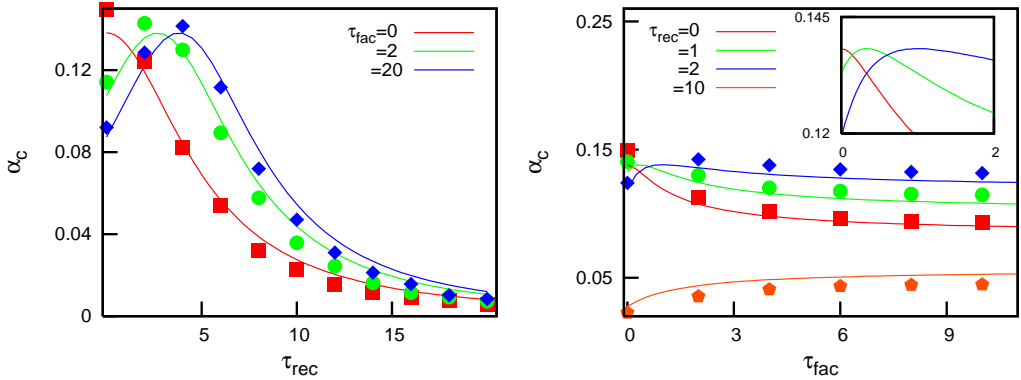,width=14cm}}
\caption{Left: $\alpha_c$ as a function of $\tau_{rec}$, for different values of $\tau_{fac}.$ Facilitation induces the appearance of non-monotonic dependences in the critical storage capacity with a maximum which reaches the limit of static synapses. Right: $\alpha_c$ as a function of $\tau_{fac},$ for different $\tau_{rec}.$ For $\tau_{rec}$ relatively small, $\alpha_c$ takes high values for the whole range of values of $\tau_{fac}.$ The inset shows a detail near the maximum of the mean field curves. Other parameters used in simulations were $U_{SE}=0.2$ and $N=3000.$}
\label{fig3}
\end{figure}
We can also obtain a non-monotonic behaviour of $\alpha_c$ if we fix $U_{SE}$ and vary the other synaptic parameters, as it is shown in Fig. \ref{fig3}.  As a function of $\tau_{rec},$ the critical storage capacity reaches the classical static limit for a certain nontrivial $\tau_{rec}$ value if facilitation is present. As a function of $\tau_{fac}$, the classical limit is also obtained providing that depression is not too strong (see, for instance, the curve for $\tau_{rec}=10$ in right panel of Fig. \ref{fig3}, where a large depression time constant induces lower $\alpha_c$ values). The figure also shows the good agreement of mean-field curves and simulations (symbols). The appearance of these maxima in Figs.~\ref{fig2}A and \ref{fig3} can be explained due to the competition between depression and facilitation mechanisms. Once the system has arrived to a fixed point of the dynamics, the effect of depression and facilitation is mainly a modification of the (fixed) strength of the synapses. Depression produces a decrease of the synaptic strength when the presynaptic neuron is active all the time. As a consequence, and compared with the static case, the pattern destabilizes for lower values of the noise produced by the interference with other patterns. This leads to a lower critical storage capacity value~\cite{bibitchkov,torresCAPACITY}. Facilitation, however, has the opposite effect as it increases the synaptic strength. Therefore, facilitation can enlarge the critical storage capacity for a given level of noise with respect to the depressing case. The increase in the synaptic strength due to facilitation can only be induced until it reaches the static synapse strength limit ---because the product $x_i F_i$ in (\ref{lfp}) cannot be larger than one. One can think that since these two mechanisms are regulated by different parameters, their competition would lead to the appearance of a maximum in $\alpha_c.$ This argument is not sufficient to explain the appearance of such a maximum since one has to consider that such competition affects both the signal term and the noise produced by the interference with other patterns. Only the consideration of the cooperative effect of all these mechanisms can explain the appearance of a maximum in $\alpha_c$, as it is observed. 

To measure how the relative strength of the signal compared with the noise is affected by the competition between these two synaptic mechanisms one can compute, for instance, the ratio between  the signal and noise contributions to the overlap (see Eq. (\ref{m})), that is
\begin{equation}
 SNR\equiv\frac{1}{1+ \left (\frac{1+\gamma\gamma'-\gamma'}{\gamma'}\right )^2,
\label{snr}
}
\end{equation}
where we used $r\approx 1$ and $m\approx 1$ which is a good approximation at $T=0.$ For static synapses one has $SNR=1.$ One can now understand the maximum appearing in figure \ref{fig2}A, for certain values of $\tau_{fac}$ and $\tau_{rec},$ as a function of $U_{SE},$ by inspection of figure  \ref{fig2}B. If one plots $SNR$ as a function of $U_{SE}$ one observes that it also has a maximum at a certain value $U^*_{SE},$ where $SNR=1,$ which is the value corresponding to the static synapse limit (in the figure this corresponds, for instance, to $U^*_{SE}\approx 0.33$ for $\tau_{rec}=2,$ $\tau_{fac}=20$). Therefore, it corresponds to the maximum observed in the behaviour of $\alpha_c$ for the same value of the synaptic parameters (see figure \ref{fig2}A). For other values of $\tau_{rec},$ $\tau_{fac}$ and $U_{SE},$ the shape of the $SNR$ should be different but, similarly to the previous example, it can easily explain the non-monotonic behaviour of $\alpha_c$ as a function of all these parameters.

\begin{figure}[t!]
\centerline{\psfig{file=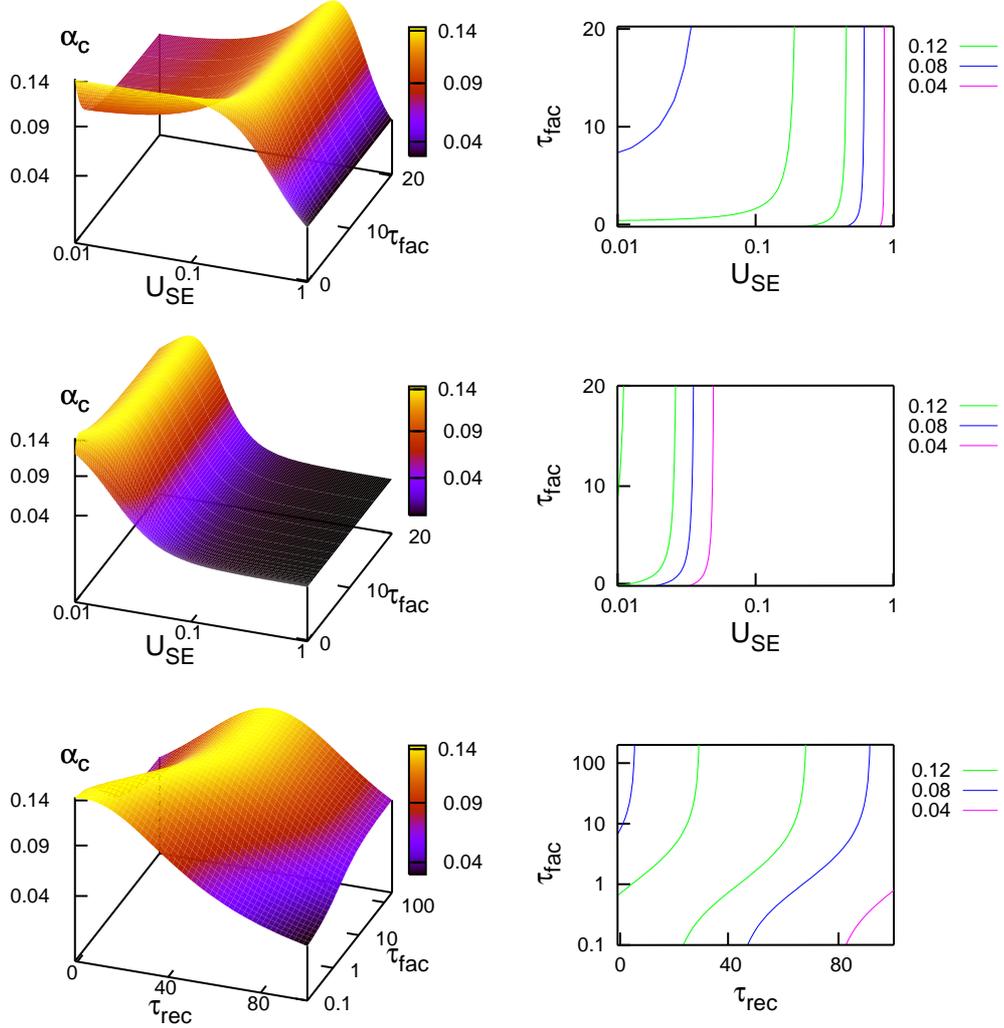,width=14cm}}
\caption{Left: Surface plots of the critical storage capacity $\alpha_c$ as a function of different synaptic parameters, for fixed $\tau_{rec}=2$ (top), $\tau_{rec}=50$ (middle) and $U_{SE}=0.02$ (bottom). Right: Contour plots which correspond to the surfaces of the left. Regions inside the lines corresponds to a set of parameters for which the critical storage capacity is high. Middle and bottom panels correspond to a more realistic set of parameters (if one assumes that the Monte Carlo step is of order of a typical refractory period of $2~ms$), and they illustrate that high capacities are obtained for a realistic values of the synaptic parameters. As one can see, the inclusion of facilitation is able to double the critical storage capacity for certain situations (see main text).}
\label{fig4}
\end{figure}

In general, we can see that large values of $\alpha_c$ appear when $U_{SE}$ and $\tau_{rec}$ have moderate values, and large enough $\tau_{fac}.$ These values coincide qualitatively well with those described in facilitating synapses of some cortical areas, where $U_{SE}$ is low compared with the corresponding values found in depressing synapses and $\tau_{rec}$ is several times lower than $\tau_{fac}$ \cite{markramPNAS}. To have such a relatively low value for $U_{SE}$ is important because it allows for a stronger recovery of the synaptic strength due to facilitation. In addition, it is worthy to note that obtaining high $\alpha_c$ values is possible for a wide range of synaptic conditions, as it is shown in right panel of Fig. \ref{fig3} and more explicitly in Fig \ref{fig4}. For instance, high capacities ($\alpha_c \geq 0.1$) can be obtained for very different values of $\tau_{fac}.$ Since actual synapses usually present a high heterogeneity in the degree of depression $\tau_{rec}$, and even more in the degree of facilitation $\tau_{fac}$ \cite{markramPNAS,markramNATURE06}, our results predict high values of $\alpha_c$ for realistic conditions. This is shown in middle and bottom panels of Fig. \ref{fig4} where the high critical storage capacity is mainly obtained for a wide range of $\tau_{fac}$ (concretely, for $\tau_{fac} > 10$) around $U_{SE}=0.02,$ and $\tau_{rec}=50$. Assuming that our Monte Carlo time step is comparable with a typical refractory period of $2~ms$, these values would correspond to $U_{SE}=0.02$, $\tau_{rec}=100~ms$ and $\tau_{fac} >20~ms$, which are within the range of realistic values in several cortical areas~\cite{markramPNAS}. Actual neural systems could, indeed, take advantage of this peculiarity to preserve a possible fine tuning of the degree of facilitation for other purposes, such as a fast dynamical processing of data,  while an optimal recall of the memories is conserved.

The improvement in the critical storage capacity for these realistic values ($U_{SE}=0.02$, $\tau_{rec}=100ms$ and $\tau_{fac} >20~ms$) in comparison with the case of only depressing synapses is highly significant. Looking at bottom panels of Fig.~\ref{fig4}, for instance, one can see that the critical storage capacity reaches $\alpha_c \simeq 0.138$ for the parameter values mentioned above. However, if we consider only the effect due to depressing synapses (that is, we set $\tau_{fac}=0$), we obtain $\alpha_c \simeq 0.07$. That is, the capacity decreases around $50{\%}$ of its value with facilitation. This indicates that facilitation could have a highly important role in the storage and recall of memories.




\begin{figure}[t!]
\centerline{\psfig{file=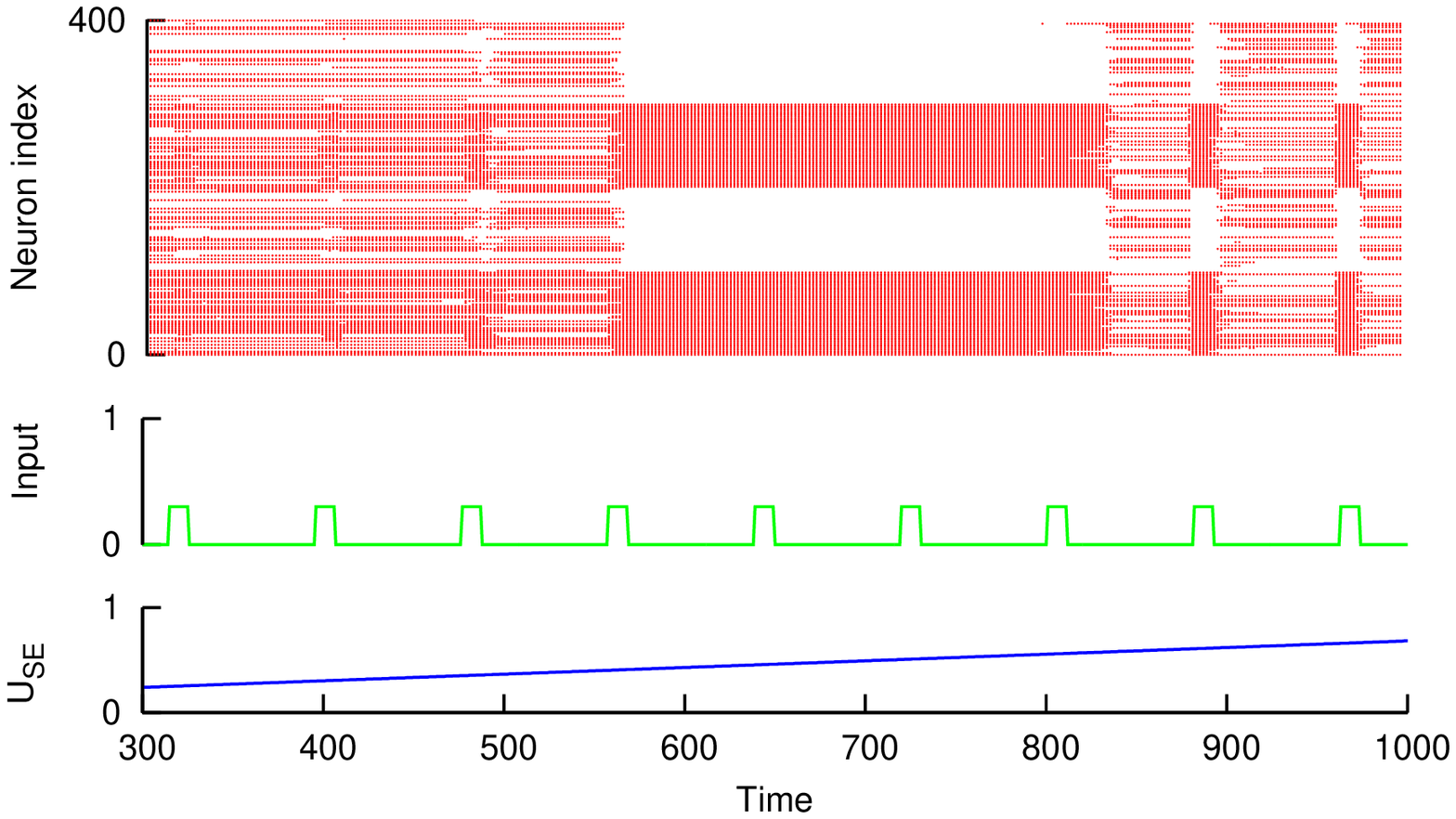,width=14cm}}
\caption{Memory retrieval under external stimulus for a network of $N=400$ neurons with dynamic synapses. The network is stimulated with a periodic weak input train (middle panel) which, in the memory phase, is able to induce the retrieval of a certain activity pattern (constructed in a band-like fashion to allow a clearer visualization). The value of $U_{SE}$ is slowly incremented in time (bottom panel) while the synaptic parameters $\tau_{rec}=2$ and $\tau_{fac}=20$ remain fixed. Good retrieval occurs only for a certain window of values of $U_{SE}$, which shows a non-monotonic  dependency of $\alpha_c(U_{SE}).$}
\label{fig5}
\end{figure}

As an illustrative example of the implications of the results reported above, let us consider a system constituted by a network of $N$ fully connected neurons which receive an additional weak external input during a very short period of time. The total synaptic current to each neuron $i$ then becomes $h_i({\bf s},t)+h_i^{ext}(t).$ If the system is in the memory phase, we expect the external input to drive the system towards a stationary attractor state $\{ \xi_i^{\mu_0} \}$. We will also consider that the input stimulus occurs periodically in time as follows
\begin{equation}
h_i^{ext}(t)=
\left\{
\begin{array}{l}
	h \xi_i^{\mu_0}~~~~~~t_n \leq t < (t_n+\delta t)\quad \forall n=0,\ldots,N_{inputs}\\\

	0 ~~~~~~~~~~~~\mbox{otherwise}
\end{array}
\right.
\end{equation}
where $h\ll 1$ is the amplitude of the weak input, $t_n$ is the time at which	the $n-$th input event occurs, $\delta t$ is the duration of a single input	event and  ${\cal T}=t_{n+1}-t_{n}$ is the period of the stimulus. If the	system is in the memory phase, the stimulus will lead the system into the	$\mu_0 -th$ attractor and the memory retrieval will be successful. Otherwise,	for large number of stored patterns, namely $M>\alpha_c N$ the system will	fall in a {\em spin-glass} state characterized by a mixture of a high number of patterns, and the retrieval will have failed. We explored how certain synapse parameters affects the retrieval process under this type of stimulus. This is shown in Fig. \ref{fig5} for a network of $N=400$ neurons and $M=48$ patterns ($\alpha =0.12$), with the pattern $\mu_0$ constituted by consecutive groups ---$100$ neurons each--- of alternate firing and silent neurons, and the remaining $M-1$ being random unbiased patterns. As an example, we consider dynamic synapses with fixed characteristic time constants $\tau_{rec}=2$ and $\tau_{fac}=20,$ and $U_{SE}$ varying in time for the whole duration of the stimulus. The figure shows that facilitation, which induces the appearance of a non-monotonic relation $\alpha_c (U_{SE}),$ allows for a good response to the external weak stimulus for a certain window of values of $U_{SE}.$ In particular, for values of $U_{SE} \simeq 0.4$ the stimulus is able to drive the system towards the attractor and recover the corresponding memory pattern $\mu_0.$ The range of values of $U_{SE}$ at which the system retrieves the pattern coincides with those between the points at which the line $\alpha=0.12$ crosses the critical mean-field line $\alpha_c (U_{SE})$ showed in Fig. \ref{fig2} for $\tau_{rec}=2$ and $\tau_{fac}=20.$ A similar type of behaviour also occurs fixing $U_{SE}$ and $\tau_{rec}$ and varying now $\tau_{fac}$ (data not shown), which also shows the main role of facilitation in memory recall.




\begin{figure}[t!]
\centerline{\psfig{file=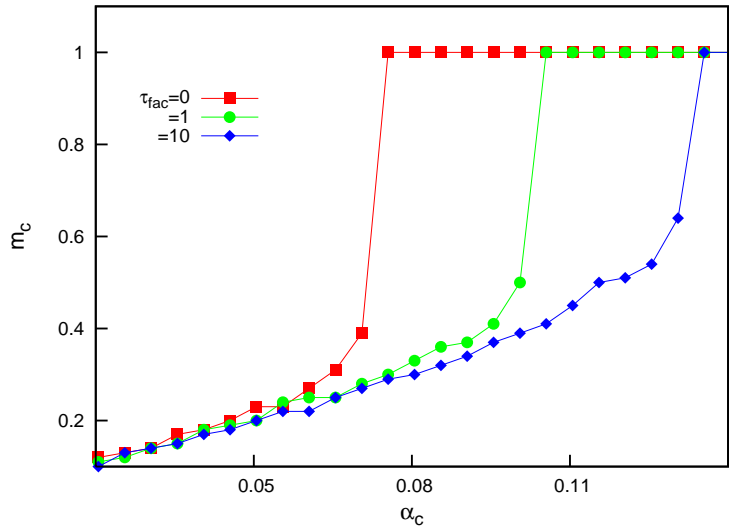,width=10cm}}
\caption{Basins of attraction of a neural network in the presence of dynamic synapses. We can see that facilitation enlarges the basins of attraction with respect to synapses with only depressing mechanisms. This allows to retrieve the previously stored patterns even if the initial condition for the network is only weakly correlated with the corresponding pattern. Parameter values are $U_{SE}=0.2$ and $\tau_{rec}=5$.}
\label{fig6}
\end{figure}

The effect caused by dynamic synapses does not only affect the stability of fixed points of the system, which gives us the critical storage capacity, but also the dynamics of the network. This has been recently investigated for the case of a single stored pattern ($\alpha\rightarrow 0$) at finite temperature~\cite{torresNC07}. For many patterns, however, the interference among them can influence the behaviour of the system near the attractors due to the appearance of many local minima associated to {\em spin-glass} states~\cite{amitAP}. To analyse this effect, we can define the basin of attraction of an activity pattern $\mu$ as the minimal value for the initial condition, $m_c^\mu$, that allows the system to tend to this attractor~\cite{bibitchkov,matsumoto07}. The measure of the basins of attraction is highly relevant because the system will not easily tend to the stored attractors if their basins are too shallow. Since we know that dynamic synapses have indeed a notable effect in the dynamics of the network at finite $\beta$, we expect them to influence the dynamics also at $\beta \rightarrow \infty$, and therefore modify the basins of attraction in some way.  We have explored this issue, and the results are shown in Fig. \ref{fig6}. We can see that the inclusion of facilitation in a network of depressing synapses leads to an increment of the basins of attraction. Since basins of attraction are associated with the error-correcting ability of the system\footnote{The error-correcting ability resembles the capacity of an attractor neural network to properly recognize an activity pattern when the given initial condition is not strongly correlated with the corresponding pattern.}, our results show that networks with facilitating synapses are adequate for recovering patterns, even from initial conditions only weakly correlated with the patterns. One can conclude, therefore, that facilitation increases the stability of the fixed points by increasing the basins, and this leads to a higher critical storage capacity, as we have reported previously.


\section{Discussion}

In this work we have focused on the role of the competition between several synaptic activity-dependent mechanisms, such as short-term depression and facilitation, in the capacity of attractor neural networks to store and retrieve information codified as activity patterns. Previous studies have found that depressing synapses drastically reduce the capacity of the network to properly retrieve patterns~\cite{bibitchkov,torresCAPACITY,matsumoto07}. These results highlight the role of depression on the processing of spatio-temporal information at short time scales (which allows for the appearance of dynamical memories), in detriment of its function in stable recall necessary for memory-oriented tasks. The consideration of additional potentiating mechanisms, such as synaptic facilitation, turns out to be convenient then for memory recall in these dynamical conditions, reaching in some cases the static limit $\alpha_c \simeq 0.138.$ This leads to think that synaptic facilitation could have a crucial role in the performance of memory retrieval tasks, while maintaining the well known nonlinear properties of dynamic synapses, convenient for information processing and coding \cite{abbott04}.

Our results also indicate that the range of parameters for which facilitation allows to have a good memory performance is notably wide, and therefore, these benefits can be achieved without a precise fine tuning of the synaptic parameters of the model. For instance, it is well known that dynamic synapses, and in particular facilitating synapses, usually present a high heterogeneity in their concrete characteristics \cite{markramPNAS,markramNATURE06}. Since the conditions for which we found high critical storage capacities for random unbiased patterns (i.e, $\alpha_c>0.1$) are very general, in the framework of our model, this can support the idea that actual neural systems could indeed take advantage of this fact to perform additional tasks---which are considerably different from a dynamical point of view--- while the optimal access to memories is maintained.

Although we have derived a mean-field theory for unbiased random patterns $f=0.5,$ there exist other mean-field approaches in the literature which can corroborate our main conclusions about storage capacity and can be useful to extend our study for other types of stored patterns. One can employ, for instance, the mean-field theory developed in~\cite{fukai93} valid also for other values of $f$, or the one presented in~\cite{tsodyksstorage88}. Our approximate theory presents, however, several differences which we consider of convenience here. Concretely, the fact that it allows to work with a network with temperature, even in an approximate way, represents a significant practical advantage. It could, in principle, be a good approximation for high temperatures, and preliminary results confirm this hypothesis \cite{mejiasT09}. In addition, the assumption of having the same threshold level for all neurons, as is done in~\cite{tsodyksstorage88}, seems to be too restrictive for the modelling of biologically motivated neural networks due to the well known variability observed in the voltage threshold of actual neurons~\cite{azouz2000}. In our work, however, this experimental fact is taken into account by considering a threshold $\theta_i$ given by (\ref{threshold}) for each neuron and which induces the existence of the noise term $B^\mu$ (which, as we have seen, has a strong effect in the behaviour of $\alpha_c.$) 

In order to treat the effect of dynamic synapses, and concretely of short-term depression and facilitation, we have employed a simple model for synapse dynamics~\cite{tsodyksNC}. The predictions of this model agree with the experimental data from cortical slices, as one can see in~\cite{markramPNAS} (see explanation in the results section). However, there are more realistic models which could be used to test our results. It is known, for example, that the stochastic nature of the transmitter release could play an important role in synaptic fluctuations~\cite{synapticnoise}. Models which take into account this stochasticity (such as~\cite{rocha05}) could be used to test our results with fluctuating synapses, although the complexity of such stochastic models would not allow to develop a simple mean-field theory, even approximate.

It is also known that dynamic thresholds are responsible for several complex phenomena in ANN~\cite{horn89}, that could be similar to the ones observed in ANN with dynamic synapses~\cite{torresNC,torresNC07}. This could lead us to think about the influence of dynamical thresholds in the network critical storage capacity, and its relation with the results presented here. While this is an interesting issue not reported yet, it is worth noting that although dynamical thresholds also induce the appearance of oscillatory states similar to the case of dynamical synapses, a direct mathematical relation between the dynamics of thresholds, as the model reported in  ~\cite{horn89} and the phenomenological model of dynamical synapses by \cite{tsodyksNC} cannot be derived (see discussion about this important issue in~\cite{torresNC}).

Finally, and attending to the dynamics and error-correcting abilities, the effect of synaptic depression on the basins of attraction has been previously studied \cite{bibitchkov,matsumoto07}. In these works, neural networks with a general inhibitory contribution are considered, and several assumptions such as a fixed threshold value for all neurons are made. On the other hand, our study considers general networks in which excitation and inhibition are treated in the same way (in particular, our neurons are not purely excitatory or inhibitory), and each neuron possesses its own particular threshold value which is also in agreement with several experimental evidences \cite{azouz2000}. Our study shows that facilitation enlarges the basins of attraction compared with the case of only depressing synapses. As a consequence, we find that a convenient balance between synaptic depression and facilitation is necessary for neural networks to work optimally at different dynamical tasks. This is in agreement with recent experimental results which show a heterogeneous level of depression and facilitation in real synapses~\cite{markramPNAS,markramNATURE06}.

\section{Acknowledgements}
This work was supported by the \textit{MEyC--FEDER} project FIS2005-00791 and the \textit{Junta de Andaluc\'{\i}a} project P06--FQM--01505. We thank S. Johnson for very useful comments.



\begin{thebibliography}{}

\bibitem[Abbott and Regehr, 2004]{abbott04}
Abbott, L.~F. and Regehr, W.~G. (2004).
\newblock Synaptic computation.
\newblock {\em Nature}, 431(7010):796--803.

\bibitem[Abbott et~al., 1997]{AVSN97}
Abbott, L.~F., Valera, J.~A., Sen, K., and Nelson, S.~B. (1997).
\newblock Synaptic depression and cortical gain control.
\newblock {\em Science}, 275(5297):220--224.

\bibitem[Amari, 1972]{amari72}
Amari, S. (1972).
\newblock Characteristics of random nets of analog neuron-like elements.
\newblock {\em IEEE Trans. Syst. Man. Cybern.}, 2:643--657.

\bibitem[Amit and Tsodyks, 1991]{tsodyksstorage91}
Amit, D. and Tsodyks, M. (1991).
\newblock Quantitative study of attractor neural networks retrieving at low
  spike rates.2. low rate retrieval in symmetrical networks.
\newblock {\em Network: Comput. Neural Syst.}, 2(3):275--294.

\bibitem[Amit et~al., 1985]{amit85}
Amit, D.~J., Gutfreund, H., and Sompolinsky, H. (1985).
\newblock Storing infinite number of patterns in a spin-glass model of neural
  networks.
\newblock {\em Phys. Rev. Let.}, 55(14):1530--1533.

\bibitem[Amit et~al., 1987a]{amitAP}
Amit, D.~J., Gutfreund, H., and Sompolinsky, H. (1987a).
\newblock Statistical mechanics of neural networks near saturation.
\newblock {\em Ann. Phys.}, 173:30--67.

\bibitem[Amit et~al., 1987b]{amitcov}
Amit, D.~J., Gutfreund, H., and Sompolinsky, H. (1987b).
\newblock Information storage in neural networks with low levels of activity.
\newblock {\em Phys. Rev. A}, 35:2293--2303.

\bibitem[Azouz and Gray, 2000]{azouz2000}
Azouz, R. and Gray, C.~M. (2000).
\newblock Dynamic spike threshold reveals a mechanism for synaptic coincidence
  detection in cortical neurons in vivo.
\newblock {\em Proc. Natl. Acad. Sci. USA}, 97(14):8110--8115.

\bibitem[Barak and Tsodyks, 2007]{tsodyks07}
Barak, O. and Tsodyks, M. (2007).
\newblock Persistent activity in neural networks with dynamic synapses.
\newblock {\em PLoS Comput. Biol.}, 3(2):323--332.

\bibitem[Bertram et~al., 1996]{bertramJNEURO}
Bertram, R., Sherman, A., and Stanley, E.~F. (1996).
\newblock Single-domain/bound calcium hypothesis of transmitter release and
  facilitation.
\newblock {\em J. Neurophysiol.}, 75(5):1919--1931.

\bibitem[Bibitchkov et~al., 2002]{bibitchkov}
Bibitchkov, D., Herrmann, J.~M., and Geisel, T. (2002).
\newblock Pattern storage and processing in attractor networks with short-time
  synaptic dynamics.
\newblock {\em Network: Comput. Neural Syst.}, 13(1):115--129.

\bibitem[de~la Rocha and Parga, 2005]{rocha05}
de~la Rocha, J. and Parga, N. (2005).
\newblock Short-term synaptic depression causes a non-monotonic response to
  correlated stimuli.
\newblock {\em J. Neurosci.}, 25 (37):8416--8431.

\bibitem[Dobrunz and Stevens, 1997]{synapticnoise}
Dobrunz, L.~E. and Stevens, C.~F. (1997).
\newblock Heterogeneity of release probability, facilitation, and depletion at
  central synapses.
\newblock {\em Neuron}, 18:995--1008.

\bibitem[Fusi and Abbott, 2007]{fusi07}
Fusi, S. and Abbott, L. (2007).
\newblock Limits on the memory storage capacity of bounded synapses.
\newblock {\em Nat. Neurosci.}, 10 (4):485--493.

\bibitem[Hebb, 1949]{hebb}
Hebb, D.~O. (1949).
\newblock {\em The Organization of Behavior: A Neuropsychological Theory}.
\newblock Wiley.

\bibitem[Hertz et~al., 1991]{hertzB}
Hertz, J., Krogh, A., and Palmer, R. (1991).
\newblock {\em Introduction to the theory of neural computation}.
\newblock Addison-Wesley.

\bibitem[Holcman and Tsodyks, 2006]{tsodyks06}
Holcman, D. and Tsodyks, M. (2006).
\newblock The emergence of up and down states in cortical networks.
\newblock {\em PLoS Comput. Biol.}, 2(3):174--181.

\bibitem[Hopfield, 1982]{hopfield}
Hopfield, J.~J. (1982).
\newblock Neural networks and physical systems with emergent collective
  computational abilities.
\newblock {\em Proc. Natl. Acad. Sci. USA}, 79(8):2554--2558.

\bibitem[Horn and Usher, 1989]{horn89}
Horn, D. and Usher, M. (1989).
\newblock Neural networks with dynamical thresholds.
\newblock {\em Phys. Rev. A}, 40:1036--1044.

\bibitem[Kamiya and Zucker, 1994]{zucker94}
Kamiya, H. and Zucker, R.~S. (1994).
\newblock Residual ca2+ and short-term synaptic plasticity.
\newblock {\em Nature}, 371(6498):603--606.

\bibitem[Markram et~al., 1998]{markramPNAS}
Markram, H., Wang, Y., and Tsodyks, M. (1998).
\newblock Differential signaling via the same axon of neocortical pyramidal
  neurons.
\newblock {\em Proc. Natl. Acad. Sci. USA}, 95(9):5323--5328.

\bibitem[Matsumoto et~al., 2007]{matsumoto07}
Matsumoto, N., Ide, D., Watanabe, M., and Okada, M. (2007).
\newblock Retrieval property of attractor network with synaptic depression.
\newblock {\em J. Phys. Soc. Japan}, 76(8):084006.

\bibitem[McGraw and Menzinger, 2003]{menzinger}
McGraw, P.~N. and Menzinger, M. (2003).
\newblock Topology and computational performance of attractor neural networks.
\newblock {\em Phys. Rev. E}, 68(4):047102.

\bibitem[Mejias and Torres, 2008]{mejiasCD07}
Mejias, J.~F. and Torres, J.~J. (2008).
\newblock The role of synaptic facilitation in spike coincidence detection.
\newblock {\em J. Comput. Neurosci.}, 24(2):222--234.

\bibitem[Mejias and Torres, 2009]{mejiasT09}
Mejias, J.~F. and Torres, J.~J. (2009).
\newblock In preparation.

\bibitem[Mongillo et~al., 2008]{mongillo08}
Mongillo, G., Barak, O., and Tsodyks, M. (2008).
\newblock Synaptic theory of working memory.
\newblock {\em Science}, 319(5869):1543--1546.

\bibitem[Pantic et~al., 2002]{torresNC}
Pantic, L., Torres, J.~J., Kappen, H.~J., and Gielen, S. C. A.~M. (2002).
\newblock Associative memory with dynamic synapses.
\newblock {\em Neural Comput.}, 14(12):2903--2923.

\bibitem[Peretto, 1992]{perettoB}
Peretto, P. (1992).
\newblock {\em An Introduction to the modeling of neural networks}.
\newblock Cambridge University Press.

\bibitem[Romani et~al., 2006]{mongillo06}
Romani, S., Amit, D., and Mongillo, G. (2006).
\newblock Mean-field analysis selective persistent activity in presence of
  short-term synaptic depression.
\newblock {\em J. Comput. Neurosci.}, 20(2):201--217.

\bibitem[Shiino and Fukai, 1993]{fukai93}
Shiino, M. and Fukai, T. (1993).
\newblock Self-consistent signal-to-noise analysis of the statistical behavior
  of analog neural networks and enhancement of the storage capacity.
\newblock {\em Phys. Rev. E}, 48(2):867--897.

\bibitem[Torres et~al., 2007]{torresNC07}
Torres, J.~J., Cortes, J., Marro, J., and Kappen, H. (2007).
\newblock Competition between synaptic depression and facilitation in attractor
  neural networks.
\newblock {\em Neural Comput.}, 19(10):2739--2755.

\bibitem[Torres et~al., 2004]{torresNEUCOM}
Torres, J.~J., Munoz, M.~A., Marro, J., and Garrido, P.~L. (2004).
\newblock Influence of topology on the performance of a neural network.
\newblock {\em Neurocomputing}, 58-60:229--234.

\bibitem[Torres et~al., 2002]{torresCAPACITY}
Torres, J.~J., Pantic, L., and Kappen, H.~J. (2002).
\newblock Storage capacity of attractor neural networks with depressing
  synapses.
\newblock {\em Phys. Rev. E.}, 66(6):061910.

\bibitem[Tsodyks and Feigelman, 1988]{tsodyksstorage88}
Tsodyks, M. and Feigelman, M. (1988).
\newblock The enhanced storage capacity in neural networks with low activity
  level.
\newblock {\em Europhys. Let.}, 6(2):101--105.

\bibitem[Tsodyks et~al., 1998]{tsodyksNC}
Tsodyks, M.~V., Pawelzik, K., and Markram, H. (1998).
\newblock Neural networks with dynamic synapses.
\newblock {\em Neural Comput.}, 10:821--835.

\bibitem[Wang et~al., 2006]{markramNATURE06}
Wang, Y., Markram, H., Goodman, P., Berger, T., Ma, J., and Goldman-Rakic, P.
  (2006).
\newblock Heterogeneity in the pyramidal network of the medial prefrontal
  cortex.
\newblock {\em Nature Neurosci.}, 9(4):534--542.

\end{thebibliography}

\end{document}